# BMC Bioinformatics

# Modular co-evolution of metabolic networks

Jing Zhao (zjane_cn@sjtu.edu.cn)
Guo-Hui Ding (gwding@sibs.ac.cn)
Lin Tao (taolin@scbit.org)
Hong Yu (yuhong@scbit.org)
Zhong-Hao Yu (damo_yu@hotmail.com)
Jian-Hua Luo (jhluo@sjtu.edu.cn)
Zhi-Wei Cao (zwcao@scbit.org)
Yi-Xue Li (yxli@sibs.ac.cn)

# Modular co-evolution of metabolic networks

Jing Zhao[1,2,4], Guo-Hui Ding[3], Lin Tao[2], Hong Yu[2], Zhong-Hao Yu[1], Jian-Hua Luo[1], Zhi-Wei Cao[2§], Yi-Xue Li [2,3,1§]

[1]School of Life Sciences & Technology, Shanghai Jiao Tong University, Shanghai 200240, China
[2]Shanghai Center for Bioinformation and Technology, Shanghai 200235, China
[3]Bioinformatics Center, Key Lab of Systems Biology, Shanghai Institutes for Biological Sciences, Chinese Academy of Sciences, Shanghai 200031, China
[4]Department of Mathematics, Logistical Engineering University, Chongqing 400016, China

[§]Corresponding author

Email addresses:
    JZ: zjane_cn@sjtu.edu.cn
    GHD: gwding@sibs.ac.cn
    LT: taolin@scbit.org
    HY: yuhong@scbit.org
    ZHY: damo_yu@hotmail.com
    JHL: jhluo@sjtu.edu.cn
    ZWC: zwcao@scbit.org
    YXL: yxli@sibs.ac.cn

# Abstract

**Background**

The architecture of biological networks has been reported to exhibit high level of modularity, and to some extent, topological modules of networks overlap with known functional modules. However, how the modular topology of the molecular network affects the evolution of its member proteins remains unclear.

**Results**

In this work, the functional and evolutionary modularity of *Homo sapiens* (*H. sapiens*) metabolic network were investigated from a topological point of view. Network decomposition shows that the metabolic network is organized in a highly modular core-periphery way, in which the core modules are tightly linked together and perform basic metabolism functions, whereas the periphery modules only interact with few modules and accomplish relatively independent and specialized functions. Moreover, over half of the modules exhibit co-evolutionary feature and belong to specific evolutionary ages. Peripheral modules tend to evolve more cohesively and faster than core modules do.

**Conclusions**

The correlation between functional, evolutionary and topological modularity suggests that the evolutionary history and functional requirements of metabolic systems have been imprinted in the architecture of metabolic networks. Such systems level analysis could demonstrate how the evolution of genes may be placed in a genome-scale network context, giving a novel perspective on molecular evolution.

# Background

Cellular functions are carried out in a modular way, and functional modules are basic building blocks of cellular organization [1]. From the perspective of molecular biology, a functional module is regarded as a group of spatially isolated or chemically specific biological components that work together for a discrete biological function. Various functional modules such as protein complexes [2-4], signalling/metabolic pathways [5-8] and transcriptional clusters [9, 10] have been detected from functional genomic techniques or bioinformatics analyses of genomic data. Recent studies suggest, to varying degrees, functional modules correlate with evolutionary modules [11], the latter being defined as cohesive evolutionary blocks in cellular systems [12, 13]. It was found that genes within functional modules tend to evolve in a coordinated way [12-15], while some fraction of evolutionary modules (or phylogenetic modules) agree well with known functional modules [16-19].

On the other hand, purely topological analysis by graph-theoretic methods has revealed that molecular networks, such as protein interaction [20-22], gene regulatory [23, 24] and metabolic networks [25-30], consist of topological modules - densely connected sub-networks within which there is a high density of edges, and between which there is a lower density of edges [31]. Since graph-theoretic methods analyze networks from topological point of view using minimal prior knowledge about biological function or evolution, they have the potential to shed new light on biological systems based on the unbiased structural information [32]. Actually, numerous studies have demonstrated, to some extent, topological modules in molecular networks tend to be functionally modularized [20-30]. Furthermore, studying molecular evolution from the viewpoint of network architecture is becoming a subject of current interest. Some recent studies suggested that the node degrees of molecular networks may constrain the evolution of proteins [33-38], and protein interaction hubs situated within modules are more evolutionarily constrained than those bridging different modules [39, 40]. However, little has been known about how network modularity affects protein evolution. Thus more studies are expected to reveal the possible correlation between topological modules and evolutionary modules in molecular networks.

In this study, we ask to which extent the identified topological modules of metabolic networks co-evolve. We explore this question by analysing the metabolic network of *H. sapiens* (hsa) reconstructed from the KEGG database [41-43]. We first break up the metabolic network into modules by the simulated annealing algorithm proposed in [28] and study the linkage pattern between modules. Then we investigate the evidence for co-evolution of modules by analysing the phylogenetic profiles, evolutionary ages and evolutionary rates of enzyme genes within modules. To mine the inherent relations between structure, function and evolution of metabolic networks, the features from *H. sapiens* network were then compared with those of the properly randomized counterparts, here, topological null model and biological null model, respectively.

## Results and discussion

### Identifying topological modules and their functions

We reconstructed the metabolic network of *H. sapiens* from the KEGG database [41-43] and represented the network by a directed substrate graph in such a way that the nodes correspond to metabolites and arcs correspond to enzyme-catalyzed reactions between these metabolites [44]. The metabolic network of *H. sapiens* consists of 1378 metabolites and 666 enzymes, with the biggest connected cluster includes 948 metabolites and 614 enzymes.

The simulated annealing algorithm [28] was utilized to decompose the metabolic network. Totally 25 topologically compact modules were obtained. In Figure 1 we display the decomposition result as a network of modules, in which each node corresponds to a module and is represented by the functional cartography [28].

Figure 1 exhibits a global view of interactions between modules, suggesting that the modules are linked in a core-periphery organized pattern [45, 46]. Some modules interact frequently and are interconnected densely to form a core, while others such as module 3, 6, 7 and 14 communicate with only one or two other modules and reside in the periphery of the network. We define the inter-module degree of one module as the number of its links with other modules, where a link by a bi-directed arc is counted as degree 2. Hence core modules have high inter-module degrees, while periphery modules have low inter-module degrees.

Further investigation revealed that most periphery modules correspond to a well-defined single pathway and perform relatively independent function. On the contrary, the core modules are mixtures of several conventional biochemical pathways, thus are difficult to be assigned a simple function. Table 1 shows that the periphery modules carry out two categories of functions: catabolism of essential human nutrients taken from the diet (essential lipids, amino acids and vitamins), biosynthesis and metabolism for complex molecules (hormones, glycans and cofactors), whereas the core modules generally perform basic metabolisms of sugars, amino acids, lipids and nucleotides. Since the *H. sapiens* metabolic network does not include biosynthetic pathways for essential human nutrients, the pathways for processing these substrates may have functioned in characterizing the species during evolutionary diversification. On the other hand, owing to the function of hormones and glycans in mediating cell recognition, communication, and signal transduction, the ability to produce and utilize these complex molecules may have led the species to develop an advanced neural system for physiological regulation along the evolutionary process [47]. Thus the functions performed by periphery modules could be regarded as some high-level metabolisms specialized for *H. sapiens*, whereas the core modules are integrated to accomplish housekeeping processes of life. Especially, the three central pathways – Embden-Meyerhof-Parnas (EMP), tricarboxylic acid (TCA) and pentose phosphate pathway (PPP) are broken up into several core modules, mainly in module 8, 11 and 13. One possible explanation is that the metabolites in these central pathways are used as common precursors for biosynthesis of universal building blocks [48] and are thus placed in different modules. In addition, these metabolites have two or more functional classifications and that they may get assigned to different modules based on features other than the central pathway they belong to. This result agrees with earlier observations concerning the high diversity of the TCA and EMP pathway [49, 50], as well as the clustering results for *Escherichia coli* (*E. coli*) metabolic network obtained by other algorithms [19, 25, 26, 30].

**The similarity between the phylogenetic profiles of enzymes within modules**

The phylogenetic profiles of any pair of enzymes can be compared to determine whether these two enzymes exhibit significant co-occurrence, so as to result in similar phylogenetic profiles. We adopted the Jaccard coefficient (JC) to measure the similarity between the phylogenetic profiles of enzyme pairs. Figure 2 (A) shows the distribution of JC for all pairwise enzymes in the *H. sapiens* metabolic network. It can be seen that most enzyme pairs exhibit low extent of similarity on their phylogenetic profiles. The average JC of the global network is 0.28. The significant level P (0.05) of the distribution is 0.66, i.e., the probability of enzyme pairs with JC bigger than 0.66 is only 0.05. Thus we set JC of 0.66 as a threshold of similarity between phylogenetic profiles and regard enzyme pairs whose JC are bigger than 0.66 as having similar phylogenetic profiles. Figure 2 (B) illustrates the relationship between the average JC for enzyme pairs within modules and the inter-module degree of module (Spearman's rank correlation is r = - 0.3814, P-value is 0.059), suggesting a statistically negative correlation tendency between these two variables. Since modules with high inter-module degrees are core modules, and those with low inter-module degrees are periphery modules, this result means that enzymes within periphery modules have higher extent of similar phylogenetic profiles than those within core modules.

We compare the similarity of phylogenetic profiles for enzyme pairs within each module with that within the global network. Figure 3(A) shows that enzyme pairs within most modules have higher average JC than those within the global network, and Figure 3 (B) demonstrates that most modules include high percentage of enzymes with similar phylogenetic profiles. We used hypergeometric cumulative distribution [51] to quantitatively measure whether a module is more enriched with enzymes of similar phylogenetic profiles (JC $\geq$ 0.66) than that by chance. Given significance level $\alpha = 0.05$, a P-value smaller than $\alpha$ demonstrates low probability that the enzymes of similar phylogenetic profiles appear in the same module by chance.

Integrating these three measures, we regard a module as an evolutionary module enriched with enzymes of similar phylogenetic profiles, if it satisfies all of the following three criteria:
(1) Average JC of the module is bigger than that of the global network.
(2) The fraction of enzyme pairs with JC $\geq$ 0.66 (definition of a threshold) in the module is significantly bigger than 0.05, i.e., the fraction of that in the global network. We set the cutoff to 0.1.
(3) The P-value is smaller than $\alpha = 0.05$.

As shown in Figure 3, all modules except module 2 satisfy the criteria (1), in which 13 modules (module 7, 3, 25, 9, 16, 4, 6, 22, 12, 15, 19, 21 and 20) also satisfy the criteria (2). Of the 13 modules, only the P-value of module 20, which equals to 0.50067, is bigger than 0.05. That is to say, although the average JC of enzyme pairs in module 20 is big enough, and this module also includes a high fraction of enzymes with similar phylogenetic profiles, this case has a high probability to occur by chance. Thus module 20 could not be regarded as an evolutionary module by our criteria.

In summary, a total of 12 modules out of 25 (module 7, 3, 25, 9, 16, 4, 6, 22, 12, 15, 19 and 21) were found to be evolutionary modules, most of which are periphery modules. The inter-modules degree of eight modules are less than 5, suggesting that periphery modules behave more cohesively in evolution than core modules. That is to say, enzyme genes within periphery modules have higher tendency to be gained/lost together than those within core modules.

**The evolutionary ages of modules**

We classified enzyme genes in the *H. sapiens* network into seven evolutionary ages: Prokaryota, Protists, Fungi, Nematodes, Arthropods, Mammalian and Human. Hypergeometric cumulative distribution [51] was used to measure whether a module is more enriched with enzymes from a particular evolutionary age than would be expected by chance. Given significance level $\alpha = 0.05$, a P-value smaller than $\alpha$ demonstrates low probability that the enzyme genes of a particular evolution age have appeared by chance.

We defined the evolutionary age of a module as the biggest value of the evolutionary age of enzymes included in this module, which satisfies all of the two criteria:
(1) More than 1/3 enzymes of this module belong to the evolutionary age;
(2) The corresponding P-value is smaller than $\alpha = 0.05$.

According to this definition, a total of 16 modules appear to belong to one specific evolutionary age. See Table 2 for the evolutionary age, percentage of enzymes in the specific age, P-value and inter-module degree for each module. As can be seen in Table 2, all of the age-1 modules except for module 12 are core modules, while the modules with later evolutionary ages are periphery modules. Figure 4 shows the statistically significant negative correlation between evolutionary age of modules and average inter-module degrees. The Spearman's rank correlation between evolutionary age and inter-module degree is r = - 0.8207 (P-value = $9.78 \times 10^{-5}$).

The distribution feature of core and periphery modules in different evolutionary ages provides some evidence for the evolutionary history of metabolic networks. Matching the evolutionary ages of modules with their functions suggests how ancient cellular functions may have evolved to gain new phenotypes with improved adaptation: the core modules appeared earlier in evolution and communicate frequently to perform the basic functions, while the periphery modules "sprout" from the compactly inter-connected core modules later via sparse linkages to carry out some novel, specialized functions. In this way, the housekeeping functions are conserved in core modules while functional specialization is achieved by extending periphery modules.

**Evolutionary rates of constituent enzyme genes of modules**

We adopted the evolutionary rate to estimate the evolutionary constraints on metabolic enzymes. A small value of evolutionary rate suggests a smaller fraction of accepted amino acid substitutions (see Methods part), hence a higher evolutionary constraint on the enzyme. We extracted the evolutionary rate of each enzyme gene included in the *H. sapiens* metabolic network from the HomoloGene database[52], computed by the approaches in [53]. Then the evolutionary rates were averaged over all enzyme genes within the same module.

The relationship between average evolutionary rate of enzyme genes within modules and inter-module degrees is shown in Figure 5. Figure 5 (A) indicates a statistically significant negative correlation between the average evolutionary rate of module and inter-module degree (Spearman's rank correlation is r = - 0.4983, P-value = 0.011.). The histogram of average evolutionary rate of modules versus binned inter-module degree is displayed in Figure 5 (B), indicating that enzymes in core modules evolve more slowly than those in periphery modules. That is to say, core modules have higher evolutionary constraints, thus are more evolutionarily conserved than periphery modules. This observation is in agreement with the housekeeping functions of core modules.

**Comparison of the *H. sapiens* network with its randomised counterparts**

In order to investigate whether the topological, functional and evolutionary features of modularity we presented above are intrinsic for metabolic networks, we constructed two versions of randomised counterparts for the *H. sapiens* metabolic network and compared them with the real network.

The first version of randomized network, called topological null model, was constructed by rewiring the links between nodes while preserving some low-level topological properties of the *H. sapiens* metabolic network. When shuffling the links, considering that the distribution of bi-directed arcs is an inherent topological feature of metabolic networks[54], we kept not only the degree of each node [55, 56], but also the total number of directed and bi-directed arcs of the metabolic network (see the algorithm in [30]). Totally 50 random networks were generated and then decomposed by simulated annealing, resulting in mean modularity metric 0.7804 and the standard deviation 0.0056. The modularity metric of the *H. sapiens* metabolic network is 0.8868, 19 multiples of standard deviation above that of the randomized counterparts, suggesting that the higher modularity of *H. sapiens* network is unlikely to arise by chance. Moreover, as shown in Figure 6 and Table A2 of additional file [see Additional file 1], the modules of the random network are so densely connected that their linkage pattern does not show a clear-cut core-periphery dichotomy. These comparisons indicate that the highly modular core-periphery organization of *H. sapiens* network could be an intrinsic topological character of metabolic networks, rather than a random phenomenon.

We generated the second version of randomized network, called biological null model, through shuffling the enzymes that catalyze the reactions while preserving the network topology. At each step of randomization, we randomly chose two reactions that are catalyzed by different enzymes and then exchanged the enzymes catalyzing them. For one randomized network, we performed the same analysis about the functional distributions, phylogenic profiles, evolutionary ages, and evolutionary rates of its topological modules as we did for the *H. sapiens* network. Although the randomized network has the same topology as the *H. sapiens* network, its topological modules are high heterogeneity of reactions from different pathways [see Figure A1 of Additional file 1], thus could not be functional specific modules. From evolutionary view, enzymes within the same modules do not exhibit similar phylogenic profiles [see Figure A2 of Additional file 1]; neither the modules have specific evolutionary ages. In addition, the correlation between the average evolutionary rates of modules and the inter-module degrees is not statistically significant (Spearman's rank correlation between evolutionary rate and inter-module degrees was calculated as $r = -0.0553$, P-value = 0.793). These results demonstrate that the topological modules of the biological null model show absence of evolutionary modularity. Therefore, the functional and evolutionary modularity of topological modules could be inherent for metabolic networks.

## Conclusions

In this study we have conducted a system-level survey about how evolution of enzyme genes is related to the structure and function of metabolic networks. From topological point of view, metabolic networks exhibit highly modular core-periphery organization pattern. Furthermore, the core modules are more evolutionarily conserved and perform some housekeeping metabolism functions, while the periphery modules appear later in evolution history and accomplish relatively specific functions. Our results suggest that the core-periphery modularity organization reflects the functional and evolutionary requirements of metabolic systems. The denser inter-connections between core modules may offer effectual protections to the basic metabolic process and keep the robustness of the metabolic system. On the other hand, the looser inter-linkages of periphery modules could function in favour of the flexibility and evolutionary ability of the system, so that the mutation or evolution of these parts may generate new phenotypes with improved adaptation while not significantly affecting other modules or even causing malfunction of the whole system. Our observation may shed light on a more global understanding of the topology, function and evolution for metabolic networks.

## Methods

### Data preparation and network reconstruction

In this study, the metabolic network of *H. sapiens* (hsa) was reconstructed using data downloaded (in May 2006) from the FTP service of KEGG (Kyoto Encyclopedia of Genes and Genomes). The "hsa_enzyme.list" file in the GENOME section of the KEGG database includes a list of the known enzymes encoded by *H. sapiens*'s genome and the corresponding genes. The "reaction" file in the LIGAND section was first scanned for all reactions catalyzed by enzymes present in *H. sapiens*'s genome, and totally 1492 reactions were determined. Then, the resulting reactions were matched to the "reaction_mapformula.lst" file, which includes direction information and the main metabolites for each reaction.

Some small molecules, such as adenosine triphosphate (ATP), adenosine diphosphate (ADP), nicotinamide adenine dinucleotide (NAD) and $CO_2$, are normally used as carriers for transferring electrons or certain functional groups and participate in many reactions, while typically not participating in product formation. Therefore, in order to reflect biologically relevant transformations of substrates, we excluded these kinds of small molecules whose list is as follows [57, 58]:
ATP, ADP, AMP, NAD, NADH, NADP, NADPH, $NH_3$, CoA, $O_2$, $CO_2$, Glu, Pyrophosphate, $H^+$.

To construct the metabolic network, substrates and products were extracted from each of the enzyme-catalyzed reactions and all resulting substrate-product pairs were listed to specify the connections between the substances. A metabolic network is represented by a directed graph whose nodes correspond to metabolites and whose arcs correspond to reactions between these metabolites, in which irreversible reactions are presented as directed arcs while reversible ones as bi-directed arcs. For example, the irreversible reaction,

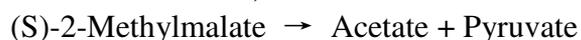
(S)-2-Methylmalate → Acetate + Pyruvate

corresponds to two directed arcs, i.e., (S)-2-Methylmalate → Acetate and (S)-2-Methylmalate → Pyruvate. The resulting metabolic network of *H. sapiens* embraces 1378 metabolites and 666 enzymes, in which 948 metabolites and 614 enzymes are included in the biggest connected cluster, while the other 52 enzymes and 430 metabolites scatter in 124 small clusters. Like previous studies of metabolic networks based on topology[25-30], we only analyze the biggest connected cluster.

### Modularity and network decomposition

In this study, we applied the simulated annealing algorithm developed by Guimera and Amaral [28] to break up the metabolic network of *H. sapiens* into modules (The software Modul-w was kindly obtained from Guimera and Amaral). This algorithm identifies topological modules by maximizing the network's modularity metric through an exhaustive search, thus may generate the best decomposition of the network.

For a given decomposition of a network, the modularity metric is defined as the gap between the fraction of arcs within clusters and the expected fraction of arcs if the arcs are wired with no structural bias [59]:

$$M = \sum_{i=1}^{r}[e_{ii} - (\sum_{j} e_{ij})^2]$$

where $r$ is the number of clusters, $e_{ij}$ is the fraction of arcs that leads between vertices of cluster $i$ and $j$. The maximum modularity metric corresponds to the partition that comprises as many as within-module links and as few as possible inter-module links.

Since the simulated annealing algorithm is stochastic, different runs may yield different decompositions of the network because of different initial conditions. The initial condition is the seed for the random number generator, which must be a negative integer. To verify the robustness of the decomposition, we first randomly generated 30 different negative integers, and then applying each of them as the initial condition to perform 30 times of independent simulated annealing and generate 30 partitions of the network. Figure 7 shows the fraction of times that each pair of nodes in the network is clustered into the same module. It is thus confirmed that the topological modules are robustly and consistently identified.

**Hypergeometric distribution and P-value**

If we randomly draw $n$ samples from a finite population, the probability of getting $i$ samples with the desired feature by chance obeys hypergeometric distribution:

$$f(i) = \frac{\binom{K}{i}\binom{N-K}{n-i}}{\binom{N}{n}},$$

where N is the size of the population, K is the number of items with the desired feature in the population. Then the probability of getting at least $k$ samples with the desired feature by chance can be represented by hypergeometric cumulative distribution defined as P-value:

$$P = 1 - \sum_{i=0}^{k-1} f(i) = 1 - \sum_{i=0}^{k-1} \frac{\binom{K}{i}\binom{N-K}{n-i}}{\binom{N}{n}}$$

Given significance level $\alpha$, which is usually set as 0.05, a P-value smaller than $\alpha$ demonstrates low probability that the items with the desired feature are chosen by chance. Hence P-value can be used to measure whether the $n$ samples drawn from the population is more enriched with items of the desired feature than would be expected by chance [51].

**Phylogenetic profiles**

For every enzyme, its phylogenetic profile is defined as a binary vector that encodes its absence (0) or presence (1) in the reference genomes. In this study, we applied the method in [16, 17] to construct the phylogenetic profiles of enzymes. We first chose 115 organisms (16 eukaryote, 83 bacteria and 16 archaea) as reference genomes. To reduce the effect of bias in the organism distribution, we then merged these 115 genomes into 54 taxa according to the NCBI taxonomy[16, 17, 52] [see Table A1 in Additional file 1]. At last, the ENZYME section of the KEGG LIGAND database [43] was utilized to determine whether the corresponding enzyme is coded in the reference genomes or not. The resulting phylogenetic profile of each enzyme is a 54-dimention binary vector.

We applied the Jaccard coefficient to measure the similarity between the phylogenetic profiles of any pair of enzymes. Jaccard coefficient is defined as follows[60],

$$JC_{ij} = \frac{n_{ij}}{n_i + n_j - n_{ij}},$$

where $n_{ij}$ is the number of taxa that encode both enzyme $i$ and $j$; $n_i, n_j$ are the number of taxa that encode enzyme $i$ and $j$, respectively.

**Evolutionary age**

Also based on the ENZYME section of the KEGG LIGAND database, we classified the enzymes in the metabolic networks into the following seven evolutionary ages according to their inferred first appearance during evolution,

1) Prokaryota: *E. coli* group [see Table A1 of Additional file 1]
2) Protists: *Plasmodium falciparum, Trypanosoma brucei, Entamoeba histolytica*
3) Fungi: *Saccharomyces cerevisiae; Schizosaccharomyces pombe*
4) Nematodes: *Caenorhabditis elegans*
5) Arthropods: *Drosophila melanogaster*
6) Mammalian: *Mus musculus; Rattus norvegicus; Canis familiaris*
7) Human: *H. sapiens*

An enzyme was assigned to evolutionary age of "Prokaryota" if an ortholog was detected in any organism of the 15[th] taxonomy in Table A1 of Additional file 1(the *E. coli* group); "Protists" if an ortholog was detected in *P. falciparum*, or *T. brucei*, or *E. histolytica* but not in *E. coli* group; "Fungi" if an ortholog was detected in *S. cerevisiae* or *S. pombe* but not in *E. coli* group, *P. falciparum, T. brucei,* and *E. histolytica,* and so forth.

**Evolutionary rate**

The evolutionary rate of an enzyme gene is defined as the normalized ratio of non-synonymous substitutions per nucleotide site ($K_a$) to synonymous substitutions per nucleotide site ($K_s$) that occurred in this enzyme gene [53]. We extracted the evolutionary rate of every enzyme gene in the *H. sapiens* metabolic network based on the complete genome of *Pan troglodytes* from HomoloGene database[52]. Nei and Gojobori's method [53] was used to calculate the synonymous and nonsynonymous substitution rates in HomoloGene database.

## Authors' contributions

JZ conceived of the study, designed the analysis, implemented the analysis and prepared the manuscript. GHD implemented part of the analysis and helped to revise the manuscript. LT, HY and ZHY helped JZ to implement the analysis. JHL managed the project. ZWC and YXL helped JZ to design the analysis, provided guidance, coordinated and participated in the biological and theoretical analyses, and revised the manuscript. All authors read and approved the final manuscript.

## Acknowledgements

We thank Dr. Luís A. Nunes Amaral and Dr. Roger Guimerà for kindly providing us the software Modul-w for network decomposition; Dr. Jingchu Luo and Dr. Mikael Huss for suggestive comments on the manuscript; and the anonymous reviewers for their helpful comments. JZ thanks Dr. Petter Holme and Dr. Mikael Huss for stimulating discussions about modularity. This work was supported in part by grants from Ministry of Science and Technology China (2006AA02Z317, 2004CB720103, 2003CB715901, 2006AA02312), National High Technology Research and Development Program of China (2006AA020805), National Natural Science Foundation of China (30500107, 30670953, 30670574)，International Cooperation Project of Science and Technology Commission of Shanghai Municipality (06RS07109), and Grant from Science and Technology Commission of Shanghai Municipality (04DZ19850, 06PJ14072，04DZ14005).

# Figure legends

**Figure 1 - Cartographic representation of the metabolic network for *H. sapiens*.**
Each circle represents a module and is coloured according to the KEGG pathway classification of the reactions belonging to it, while the arcs reflect the connection between clusters. The area of each colour in one circle is proportional to the number of reactions that belong to the corresponding metabolism. The width of an arc is proportional to the number of reactions between the two corresponding modules. For simplicity, bi-directed arcs are presented by grey edges. The modularity metric of this decomposition is 0.8868.

**Figure 2 The distribution of Jaccard coefficient (JC) for enzyme pairs in the global metabolic network and its modules.**

**(A) The distribution of JC for all pairwise enzymes in the *H. sapiens* metabolic network.**

**(B) The relationship between the average JC for enzyme pairs in modules and the inter-module degree of module**

**Figure 3 Comparison of the similar extent of phylogenetic profiles for enzyme pairs within each module with that within the global metabolic network of *H. sapiens.***

**(A) Average JC of enzyme pairs within modules.**

The red column represents the global network. The modules are ordered according to their average JC in a decreasing way.

**(B) Percentage of enzyme pairs within modules with JC $\geq$ 0.66 (threshold definition).**

The red column represents the global network. The modules are drawn in the same order as in (A).

**Figure 4 Relationship between evolutionary age of modules and average inter-module degrees**

**Figure 5 Relationship between average evolutionary rate of enzyme genes within modules and inter-module degrees**
(A) The average evolutionary rate of module is negative correlation with the inter-module degree
(B) Core modules (inter-module degree > 5) are evolutionarily constrained at higher extent than periphery modules (inter-module degree $\leq$ 5) do.

**Figure 6 Decomposition of one randomized network of the *H. sapiens* network, shown as a network of modules.**
Each circle represents a module, while the arcs reflect the connection between clusters. The width of an arc is proportional to the number of links between the two corresponding modules. For simplicity, bi-directed arcs are presented by grey edges. The modularity metric of this decomposition is 0.7845.

**Figure 7 Accuracy of simulated annealing algorithm to identify topological modules.**
We obtain 30 independent decompositions of the *H. sapiens* metabolic network and plot the fraction of times that each pair of nodes is clustered into the same module.

# Tables

**Table 1 Inter-module degrees (d), number of nodes (N), and main functions of the topological modules for *H. sapiens* network**

| Module | d | N | *Function category* and Main Function |
|---|---|---|---|
| 6 | 1 | 38 | *Metabolism of essential human nutrients:* Arachidonic acid metabolism |
| 24 | 1 | 41 | *Metabolism of essential human nutrients:* Phenylalanine, Tyrosine metabolism |
| 17 | 3 | 19 | *Metabolism of essential human nutrients:* Valine, leucine and isoleucine degradation, Propanoate metabolism |
| 12 | 4 | 20 | *Metabolism of essential human nutrients:* One carbon pool by folate |
| 21 | 7 | 21 | *Metabolism of essential human nutrients:* Nicotinate and nicotinamide metabolism |
| 19 | 1 | 34 | *Hormonal compound biosynthesis and metabolism:* Androgen and estrogen metabolism |
| 25 | 1 | 34 | *Hormonal compound biosynthesis and metabolism:* C21-Steroid hormone metabolism |
| 22 | 2 | 20 | *Hormonal compound biosynthesis and metabolism:* Biosynthesis of steroids |
| 20 | 5 | 28 | *Hormonal compound biosynthesis and metabolism:* Fatty acid biosynthesis |
| 7 | 2 | 26 | *Glycan biosynthesis and metabolism:* Blood group glycolipid biosynthesis - neo-lactoseries |
| 16 | 2 | 43 | *Glycan biosynthesis and metabolism:* N-Glycan biosynthesis; Fructose and mannose metabolism |
| 3 | 4 | 30 | *Glycan biosynthesis and metabolism:* Blood group glycolipid biosynthesis- neo-lactoseries; Ganglioside biosynthesis |
| 1 | 5 | 23 | *Glycan biosynthesis and metabolism:* Glycosphingolipid metabolism |
| 5 | 5 | 26 | *Glycan biosynthesis and metabolism:* Aminosugars metabolism |
| 14 | 1 | 17 | *Metabolism of cofactor:* Hemoglobin and urobilinogen biosynthesis |
| 4 | 8 | 41 | * |
| 15 | 11 | 24 | * |
| 2 | 13 | 47 | * |
| 23 | 14 | 56 | * |
| 9 | 17 | 42 | * |
| 18 | 17 | 46 | * |
| 10 | 21 | 58 | * |
| 8 | 22 | 66 | * |
| 13 | 25 | 57 | * |
| 11 | 30 | 91 | * |

* represents that the corresponding module includes several pathways thus is difficult to assign it a simplex function.

**Table 2 Evolutionary ages of topological modules for *H. sapiens* network**

| Evolutionary age | Module | Percentage of enzymes in this age | P-value | Inter-module degree |
|---|---|---|---|---|
| 6 | 3 | 82.35 % | $9.26 \times 10^{-8}$ | 4 |
| 6 | 6 | 68.75 % | $2.46 \times 10^{-4}$ | 1 |
| 6 | 7 | 71.43 % | $6.19 \times 10^{-3}$ | 2 |
| 6 | 19 | 63.16 % | $3.02 \times 10^{-4}$ | 1 |
| 6 | 25 | 85.71 % | $1.58 \times 10^{-6}$ | 1 |
| 4 | 24 | 30.00 % | $2.22 \times 10^{-3}$ | 1 |
| 3 | 20 | 40.00 % | $4.18 \times 10^{-2}$ | 5 |
| 3 | 22 | 33.33 % | $2.34 \times 10^{-3}$ | 2 |
| 2 | 1 | 36.84 % | $1.89 \times 10^{-2}$ | 5 |
| 2 | 5 | 40.00 % | $6.82 \times 10^{-3}$ | 5 |
| 1 | 8 | 69.09 % | $4.27 \times 10^{-5}$ | 22 |
| 1 | 9 | 84.38 % | $3.5 \times 10^{-6}$ | 17 |
| 1 | 12 | 65.22 % | $2.53 \times 10^{-2}$ | 4 |
| 1 | 13 | 65.31 % | $6.48 \times 10^{-4}$ | 25 |
| 1 | 15 | 64.00 % | $2.65 \times 10^{-2}$ | 11 |
| 1 | 18 | 68.57 % | $1.61 \times 10^{-3}$ | 17 |

Evolutionary age is defined as: 1. Prokaryota; 2. Protists; 3.Fungi; 4.Nematodes; 5. Arthropods; 6. Mammalian; 7.Human. See Method part for the detailed definition.

## Additional files

Additional file 1
File format: PDF
Title: Addition File for "Modular co-evolution of metabolic networks"
Description: Supplementary material for this paper

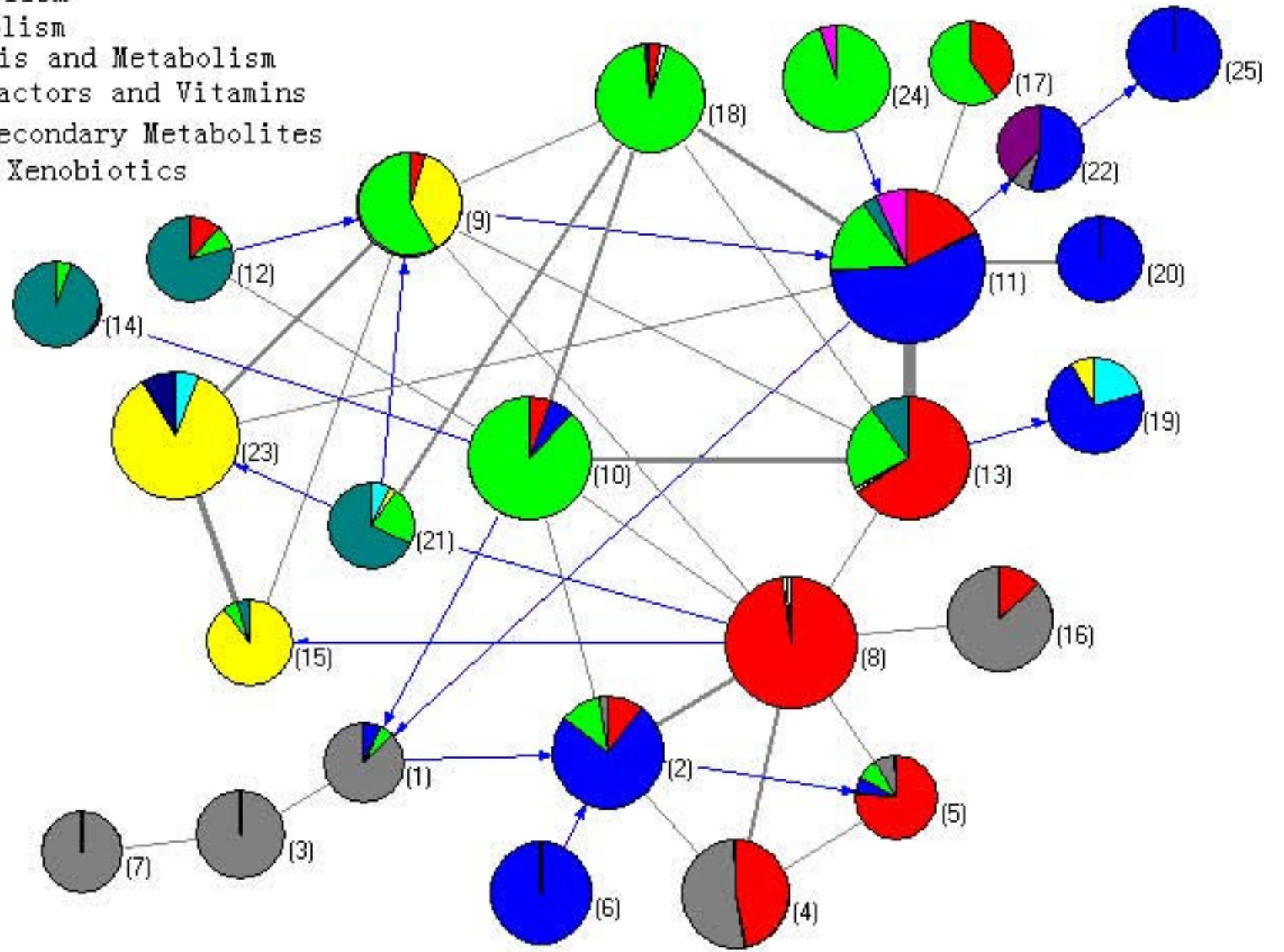

Figure 1

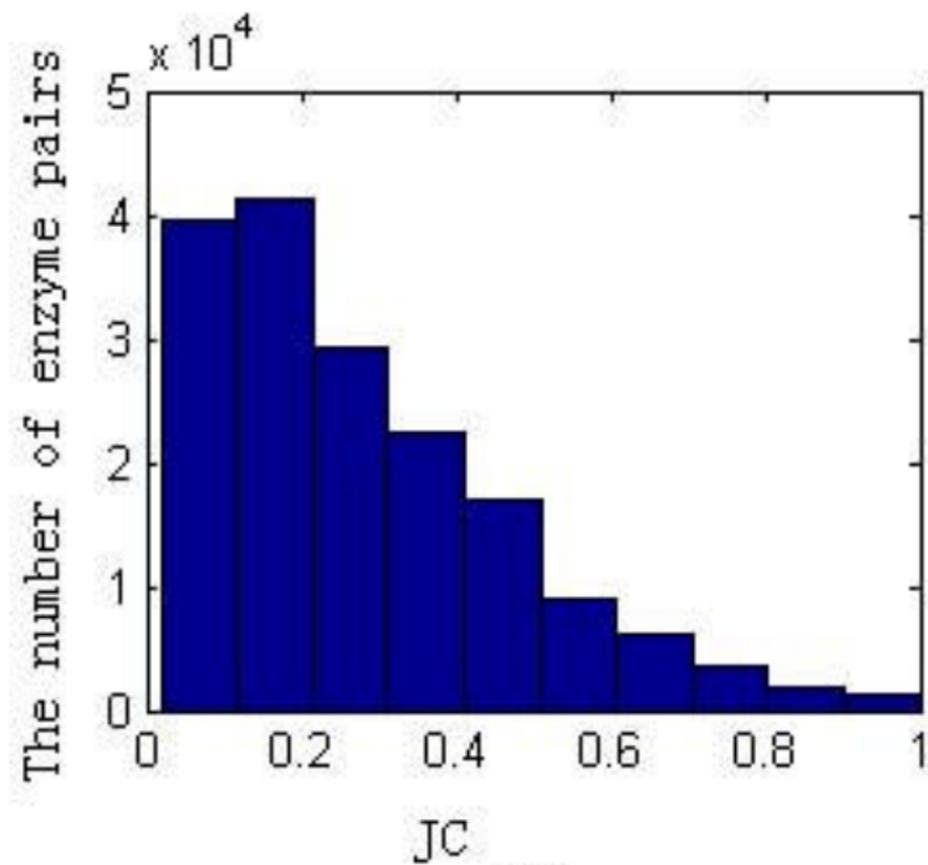 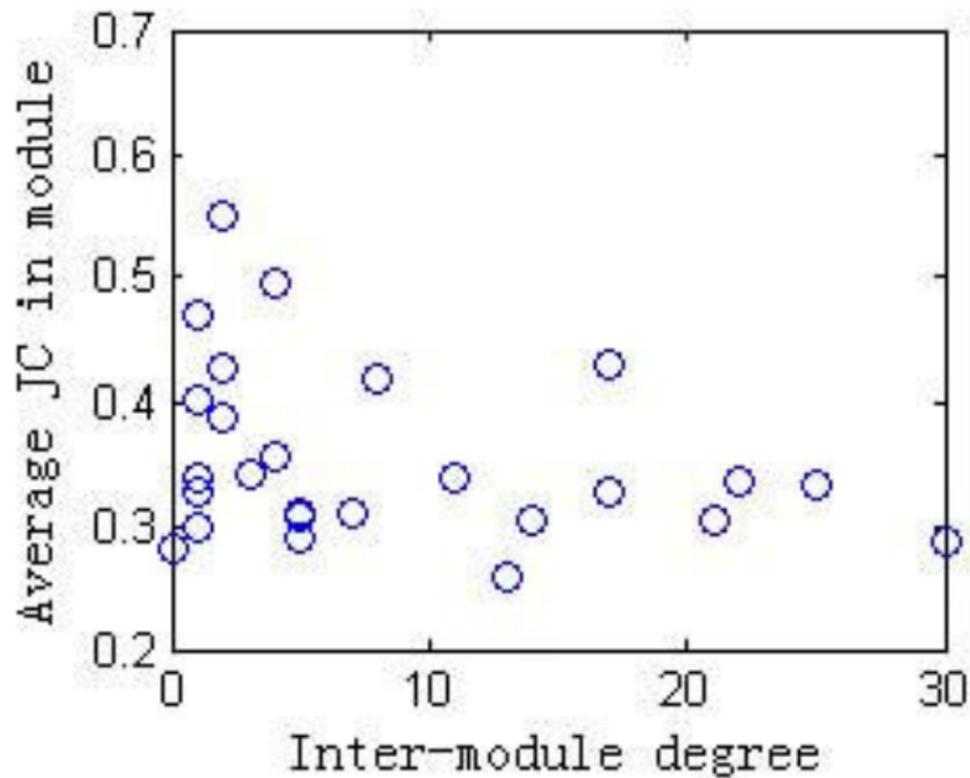

Figure 2

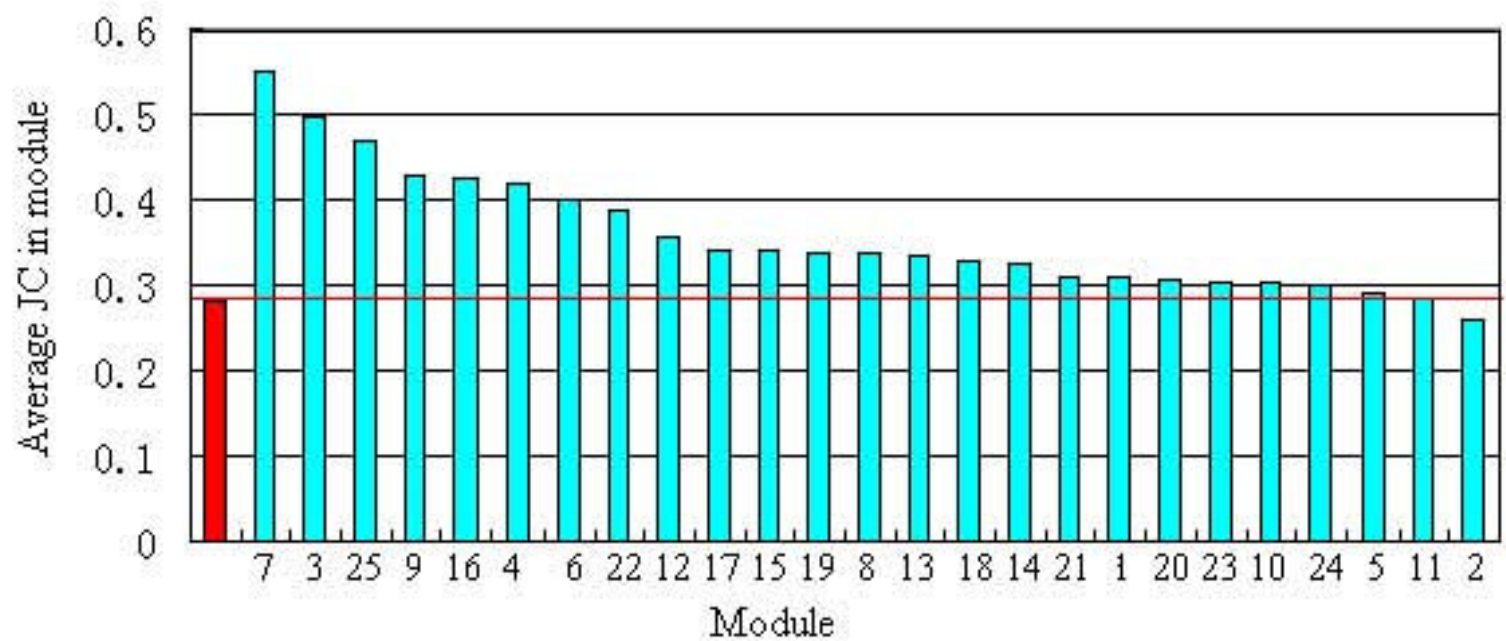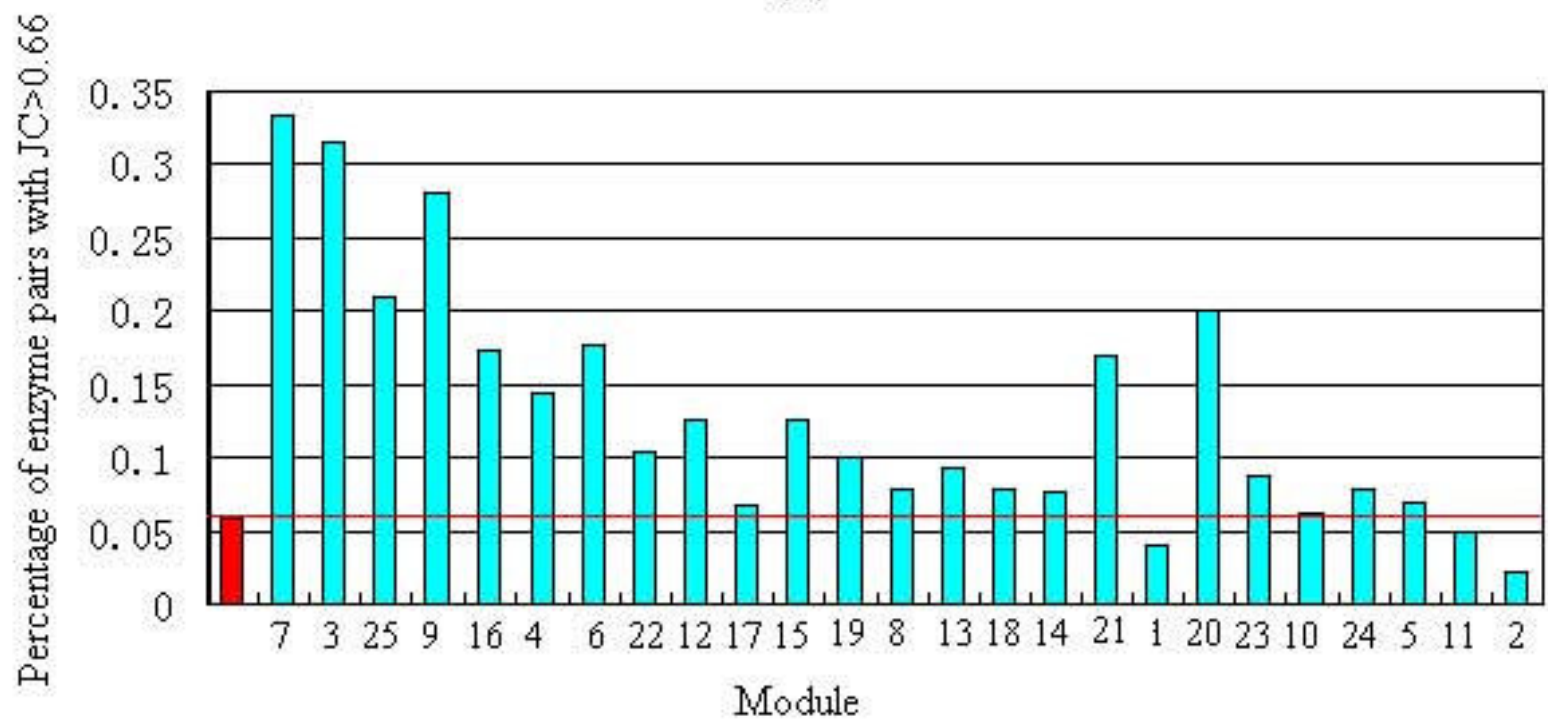

Figure 3

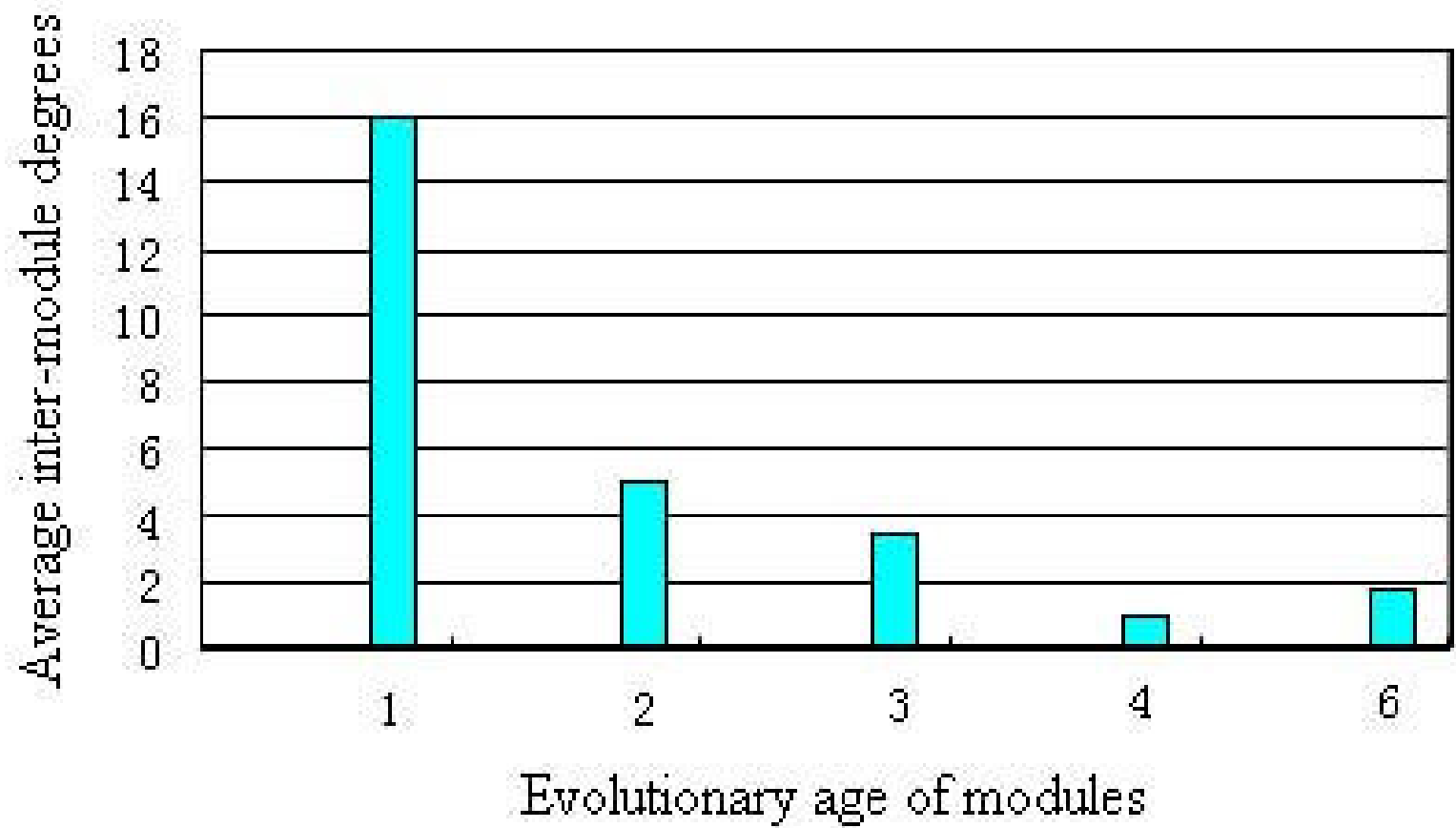

Figure 4

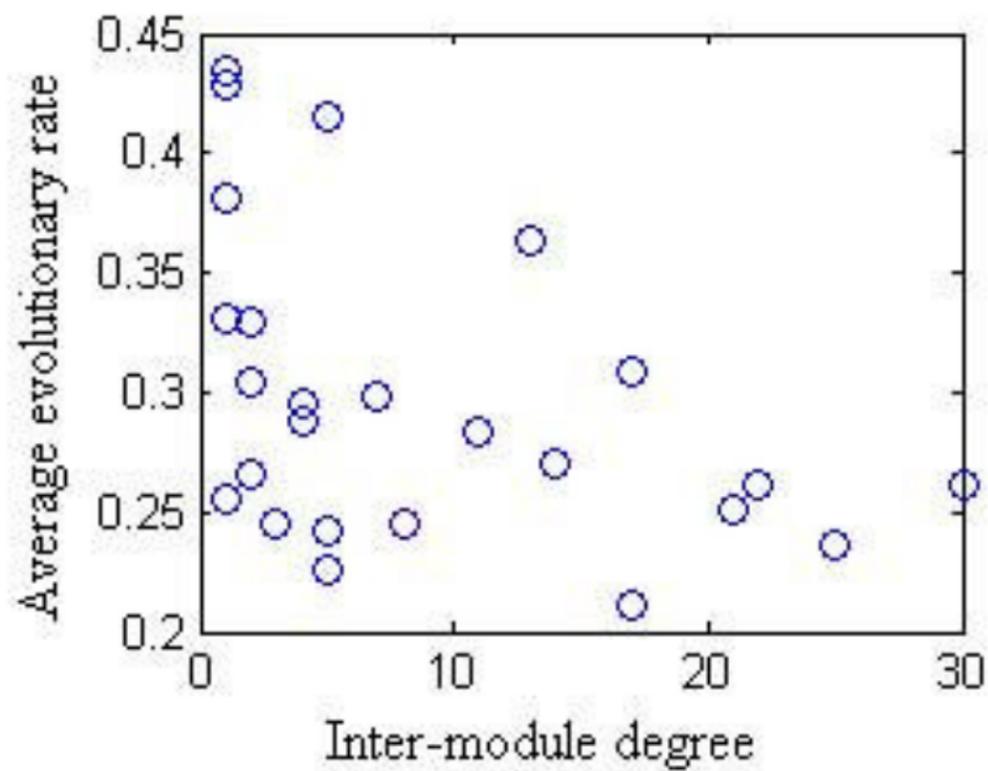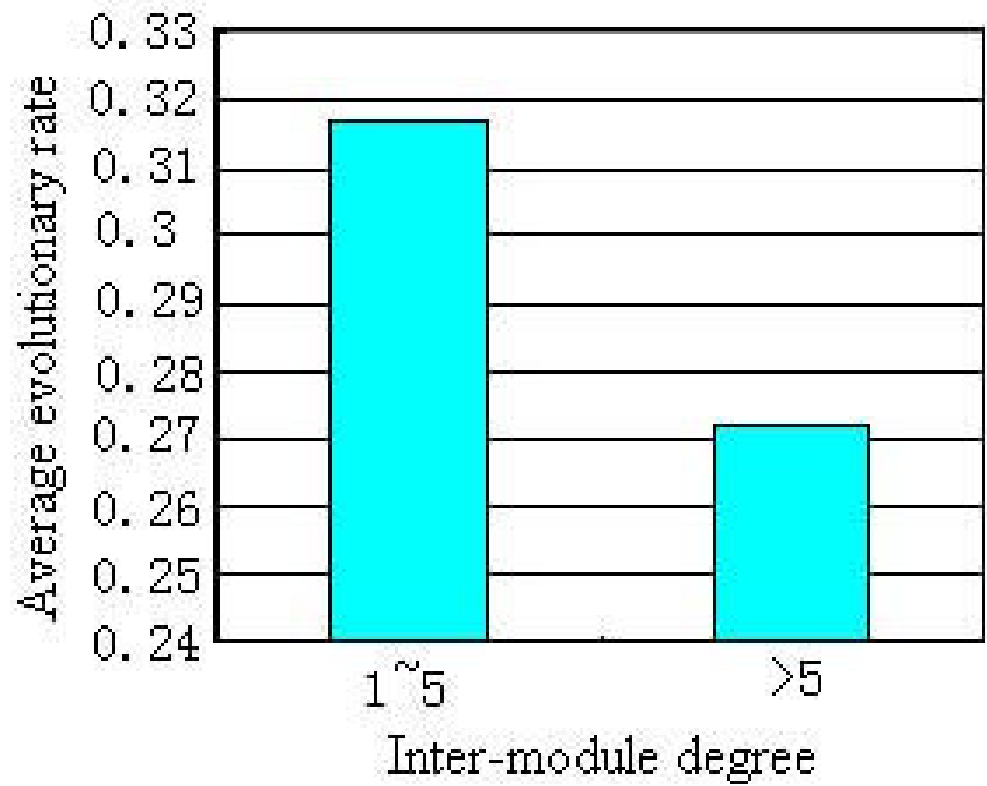

Figure 5

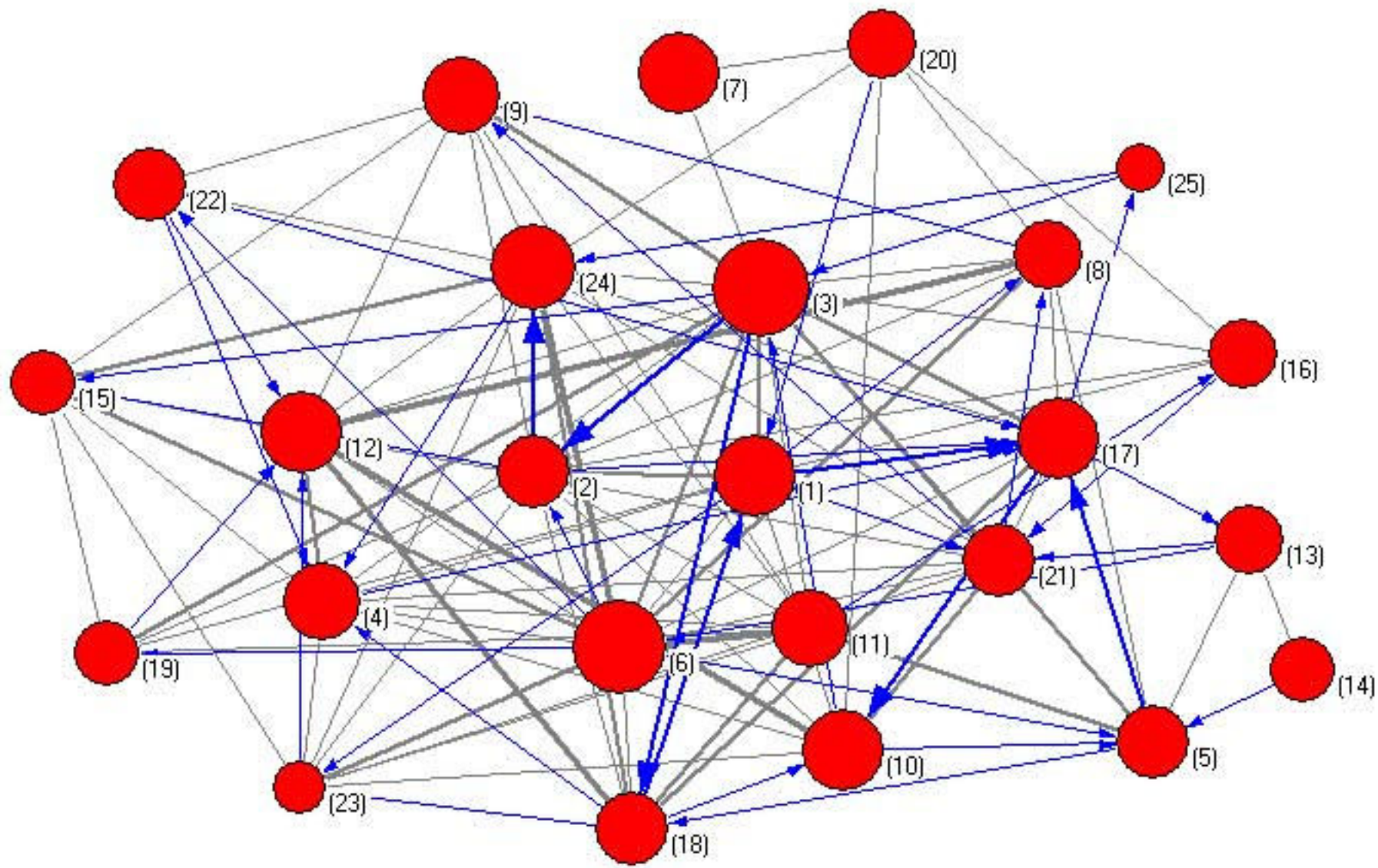

Figure 6

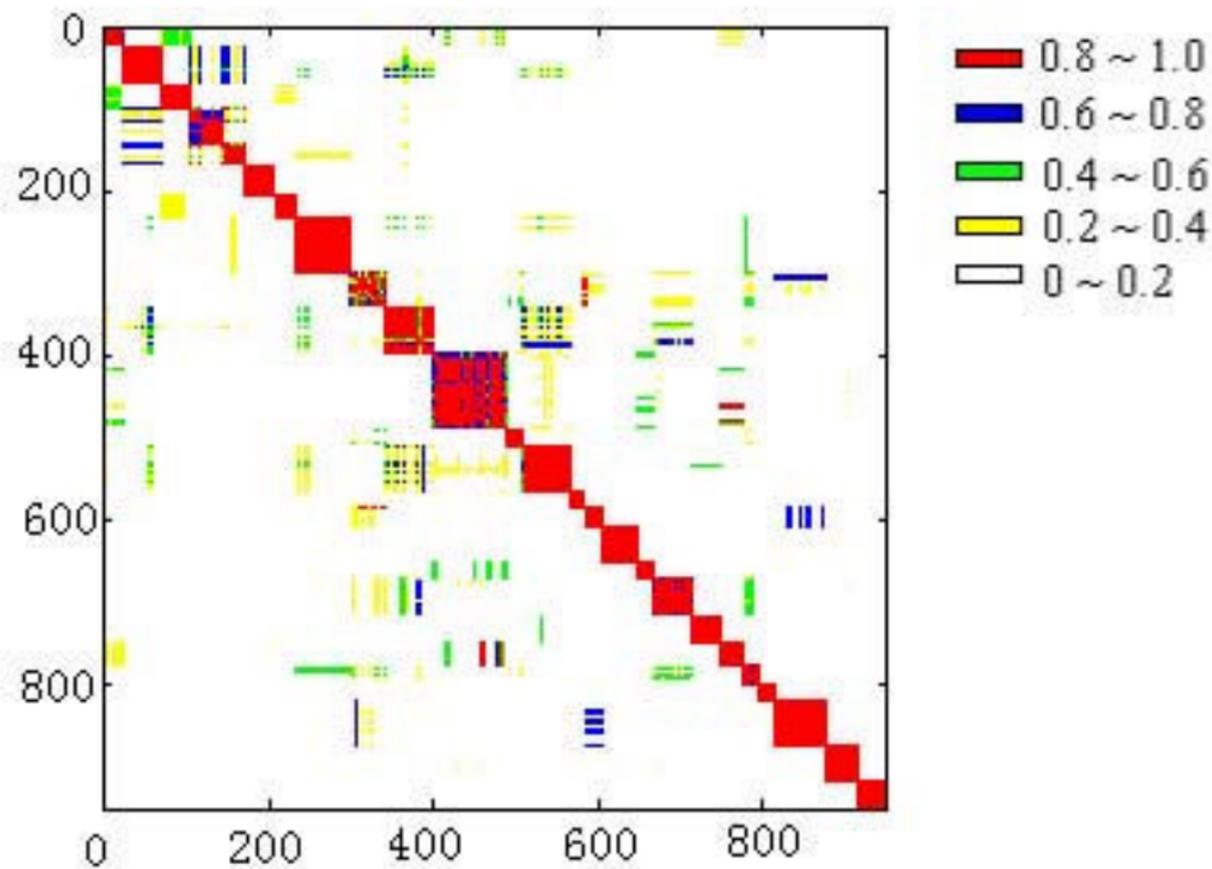

Figure 7

**Additional files provided with this submission:**

Additional file 1: modularcoevolution_additional file_070601.pdf, 79K
http://www.biomedcentral.com/imedia/1513728159146399/supp1.pdf